\documentclass[prl, aps, twocolumn, superscriptaddress,nofootinbib,preprintnumbers,
longbibliography, notitlepage, 10pt]{revtex4-1}
\pdfoutput=1
\usepackage{amsmath, latexsym, amssymb, graphicx, ifpdf, slashed, color, hyperref, 
url, cancel, bbm, hyperref}
\usepackage{comment}
\usepackage[T1]{fontenc}
\usepackage[dvipsnames]{xcolor}
\hypersetup{colorlinks, citecolor=nicegreen, linkcolor=nicered, urlcolor=niceblue}
\definecolor{nicered}{rgb}{.7,.1,.1}
\definecolor{nicegreen}{rgb}{.2,.7,.1}
\definecolor{niceblue}{rgb}{0.1,0.2,0.6}
\definecolor{darkblue}{rgb}{0,0,.5}
\usepackage{orcidlink}
\usepackage{notes2bib}

\begin{document}

\title{Blast-frozen Dark Matter and Modulated Density Perturbations}

\author{Miha Nemev\v{s}ek \orcidlink{0000-0003-1110-342X} }
\email{miha.nemevsek@ijs.si}
\affiliation{Faculty of Mathematics and Physics, University of Ljubljana, 
Jadranska 19, 1000 Ljubljana, Slovenia}
\affiliation{Jo\v{z}ef Stefan Institute, Jamova 39, 1000 Ljubljana, Slovenia}

\author{Yue Zhang \orcidlink{0000-0002-1984-7450} }
\email{yzhang@physics.carleton.ca}
\affiliation{Department of Physics, Carleton University, Ottawa, ON K1S 5B6, Canada}

\date{\today}

\vspace{1cm}

\begin{abstract}
First-order phase transitions (FOPT) are ubiquitous in beyond the Standard Model 
physics and leave distinctive echoes in the history of early universe. 
We consider a FOPT serving the well-motivated role of dark matter mass generation 
and present {\it blast-frozen dark matter} (BFDM), which transitions from radiation 
to non-relativistic relic in a period much shorter than the corresponding Hubble time. 
Its cosmological imprint are strong oscillations in the dark matter density
perturbations that seed structure formation on large and small scales. 
For a FOPT occurring not long before the matter-radiation equality, next generation
cosmological surveys bear a strong potential to discover BFDM and in turn establish
the origin of dark matter mass.
\end{abstract}

\preprint{}

\maketitle

%
%  Introduction
%
{\bf Introduction.}
At present, the $\Lambda$CDM model remains the leading candidate to explain most of 
the cosmological observations.
At zeroth order, dark matter (DM) needs to be significantly abundant, comprising 
about a quarter of the critical density today~\cite{Kolb:1990vq}, and an even higher 
fraction early on during the formation of large scale structure (LSS) of the universe.
At first order, observational cosmology has provided a wealth of knowledge about the 
DM density perturbations encoded in the matter power spectrum~\cite{Reid:2009xm, BOSS:2012tuc}.
This is further processed by the baryon-photon fluid in the early universe and manifests
itself as angular anisotropies of the comic microwave background~\cite{Tegmark:2002cy, Planck:2018vyg}.
The state-of-the-art measurements of the $\Lambda$CDM parameters have reached 
percent-level~\cite{Turner:2022gvw}.
Even higher level of precision is anticipated from cosmological observations in the 
upcoming decade~\cite{LSSTDarkEnergyScience:2018yem, Euclid:2019clj, Euclid:2021qvm,
Sailer:2021yzm}.

The key lesson from the exploration of cosmological data is that DM must be 
sufficiently cold, regardless of its identity.
If populated through thermal contact with the early universe Standard Model 
plasma~\cite{Lee:1977ua, Dodelson:1993je, Hall:2009bx, DeGouvea:2019wpf, An:2023mkf}, DM has to experience a stage 
of cooling to turn non-relativistic, either through the Hubble expansion or entropy 
dilution~\cite{Scherrer:1984fd, Nemevsek:2022anh}.
An equally important and likely related puzzle is how and when DM acquired its mass.
While the $\Lambda$CDM defines DM to be massive and cold throughout the
history of the universe, this is only necessary since the formation of the smallest 
observed structures~\cite{Viel:2013fqw, Gilman:2019nap, DES:2020fxi}.
This leaves open possibilities for exploring the origin of DM mass~\cite{Casas:1991ky, Anderson:1997un, Das:2018ons, Davoudiasl:2019xeb, Mandal:2022yym,
Das:2023enn, Chakraborty:2024pxy} and its coldness, 
as well as for hunting for the associated signatures in forthcoming experiments.

In this Letter we explore ramifications of a FOPT in the dark sector that contributes 
to the spontaneous generation of DM mass.
A FOPT proceeds through the percolation of bubbles and may complete within a period 
much shorter than the corresponding Hubble time.
We assume DM is nearly massless at early times and obtains most of its mass 
through the FOPT, which occurs before matter-radiation equality (MRE).
For a sufficiently large mass, compared to the temperature in the new phase, the 
DM fluid transitions from radiation to matter: its equation of state $w = p/\rho$
goes from 1/3 to approximately 0.
The FOPT thus acts not only as the source of DM mass, but also as a cosmic blast freezer.

How can we discover such a scenario with cosmological observables?
Bubble collisions during FOPT produce a stochastic gravitational wave (GW) background.
A number of existing and future GW detectors are sensitive to FOPT at 
$\sim 10$ MeV temperatures and above~\cite{Ballmer:2022uxx, Caprini:2024hue, Thrane:2013oya, 
GWplots}.
Recently,~\cite{Liu:2022lvz, Elor:2023xbz, Buckley:2024nen} argued that a FOPT could 
result in detectable curvature or isocurvature perturbations.
All of them are tied to bubble physics.

We point out a smoking-gun signal of the mass generating FOPT in the DM
density perturbations in the space away from the expanding bubbles.
Abruptly changing the dynamics of DM, characterized by $w$, produces a novel
oscillating effect in the matter power spectrum $P(k)$.
It applies to modes entering the horizon prior to the FOPT and is controlled by 
two phase transition parameters, the nucleation temperature $T_*$ and the inverse
duration of the PT $\beta$, together with the blast-frozen fraction $f_{\text{BF}}$ 
of all the DM in the universe.
In the large $\beta$ limit we derive an analytic expression for the modulated matter
perturbations that is valid for all $k$ modes and agrees well with the numerics.

%
%  Growing bubbles of cold dark matter 
%
{\bf DM equation of state during FOPT.}
The universe undergoes a FOPT below a critical temperature $T_c$ at which two vacua 
become degenerate.
A new phase of the universe is born through bubble nucleation.
The probability of nucleating a bubble, i.e. the false vacuum decay rate per unit
volume, is given by $\gamma = \Gamma/V = A \, e^{-B}$.
The prefactor is roughly given by $A \sim T^4$, and the exponent is $B \simeq S_3(T)/T$, 
where $S_3$ is the 3D Euclidean action~\cite{Affleck:1980ac, Linde:1981zj}.
As the FOPT proceeds, bubble nucleation starts through a sharp drop in the $S_3(T)$, 
which lifts the exponential suppression.
In other words, the time (temperature) dependence of $\gamma$ through the exponential 
dominates over the power-law time dependence of the prefactor.
The inverse duration of the FOPT is set by
$\beta \simeq -\text{d}B(t_*)/\text{d}t$~\cite{Enqvist:1991xw, Grojean:2006bp, Caprini:2019egz}, 
where $t_*$ marks the onset of efficient bubble nucleation.

The volume fraction of the universe in the false vacuum is then given
by~\cite{Guth:1982pn}
\begin{align}
  {\cal F}(t) &= \exp \left( - \int_{t_c}^t \text{d} t_1 \gamma(t_1) a(t_1)^3 
  V(t_1, t) \right) \, ,
\end{align}
where $a$ is the scale factor of the universe. 
At time $t$, the volume of a bubble born at an earlier time $t_1 < t$ is given by 
$V(t_1, t) = 4 \pi R^3/3$.
In a radiation dominated universe, the bubble radius is approximated by 
$R \simeq v_w (a(t) - a(t_1))/(H_* a_*^2)$, 
where $v_w$ is the bubble wall velocity and $a_*$ is the scale factor at $t_*$, with $H_*$ 
being the corresponding Hubble parameter.
Defining the nucleation time with ${\cal F}(t_*)=1/e$, the fractional volume is 
approximately ${\cal F}(t) \simeq \exp \left[ - \exp \beta (t - t_*) \right]$,
as derived in End Matter~\hyperref[ed1]{I} and~\cite{Enqvist:1991xw}.
The time dependence of the DM equation of state can be well modeled by
\begin{align}\label{eq:modelw}
  w(t) &\simeq \frac{1}{3} {\cal F}(t) \simeq \frac{1}{3} \exp \left[ - e^{\beta( t - t_*)
  } \Theta(t - t_c) \right] \, ,
\end{align}
where $\Theta$ is a unit-step function.
%
%  (we neglect the remaining damped oscillations and the remaining $T/m$ scaling of cold DM). 
%
The equation of state $w(t)$ is approximately continuous at $t_c$, as long as 
$\beta (t_*-t_c)\sim \beta/H_* \gg 1$.

Cranking up $\beta/H_* \gg 1$ corresponds to a nearly instantaneous FOPT. 
We derive a closed form solution for DM density perturbations in this limit, where most of 
the details of the FOPT, such as the exact value of $\beta$ and bubble wall velocity $v_w$, 
become irrelevant.
The key properties of the equation of state are
\begin{align} \label{eq:genericw}
  w(t) &= \begin{cases}
  \frac{1}{3} \, , & t < t_- \, ,
  \\
  0 \, , & t > t_+ \, ,
  \end{cases}
  & \dot w(t_-) &= \dot w(t_+) = 0 \, ,
\end{align}
where $t_\mp$ are the times immediately before/after the FOPT. 
In the $\beta/H_* \to \infty$ limit, $t_\mp \to t_*$.
We neglect any spatial dependence leading to additional gradient terms in DM perturbation 
equations, i.e. $w(t, \vec x) \to w(t)$.

{\bf Insta-freeze perturbations.}
Let us calculate the linear growth of BFDM density perturbations in the presence of 
a mass-generating FOPT. 
In the conformal Newtonian gauge, the perturbed FRW metric is
\begin{equation}
  \text{d} s^2 = a \left( \tau \right)^2 \left[ -\left(1 + 2 \psi \right) 
  \text{d}\tau^2 + \left(1 - 2 \phi \right) \text{d} \vec{x}\cdot \text{d} \vec{x} \right] \, ,
\end{equation}
where $\tau$ is the conformal time $\text{d} t = a(\tau) \text{d} \tau$ and 
$\psi, \phi$ are space-time dependent perturbations.
Prior to MRE, we have the approximate relation 
$\tau \simeq (\text{d}a/\text{d}\tau/a)^{-1} = 2 (\sqrt{a+a_{eq}} - \sqrt{a_{eq}})/(
H_0 \sqrt{\Omega_m})$, where $a_{eq} = \Omega_r/\Omega_m$ is the scale factor at MRE, 
$\Omega_r = 9\times10^{-5}$ and $\Omega_m = 0.315$ are the energy density fractions of 
radiation and matter in the universe today, and $H_0$ is the Hubble constant.

In momentum space, the DM density perturbations satisfy two linear equations~\cite{Ma:1995ey}
\begin{align}\label{eq:DMPert1}
  \delta' &= - \left(1 + w \right) \left( \theta - 3\phi' \right) - 
  \frac{3a'}{a} \left( \frac{\delta p}{\delta \rho} - w \right) \delta \, , 
  \\ \nonumber
  \!\!\!\! \theta' &= - \frac{a'}{a}(1-3w) \theta - \frac{w'}{1+w}
  \theta + \frac{\delta p/\delta \rho}{1+w} k^2\delta + k^2 \psi \, ,
\end{align}
where $\delta$ is the DM energy density fluctuation, $\theta$ is the divergence of
the DM fluid velocity, $^\prime$ stands for $\text{d}/\text{d}\tau$, and $k$ is the
co-moving momentum.
We neglect small anisotropic stress perturbations in the energy-momentum tensor
by setting $\psi = \phi$.

In radiation dominated universe, one can solve the Einstein equations to 
obtain~\cite{Dodelson:2003ft} the metric perturbation
\begin{align} \label{eq:Phiktau}
  \phi \left(k,\tau \right) = 2 \mathcal{R}(k) \, \frac{ \sin x - x \cos x}{x^3}  \, ,
\end{align}
where $x \equiv k \tau/\sqrt 3$ and $\mathcal{R}(k)$ isa nearly scale-invariant primordial curvature
perturbation produced by inflation, which has completed early on.
In the superhorizon limit $x \to 0$ and the initial condition for the gravitational
potential is $\phi(k, 0) = 2\mathcal{R}(k)/3$.

One can solve the DM perturbations analytically with a constant 
$w = \delta p/\delta \rho = 1/3$ and 0 in the two separate $\tau < \tau_-$ and 
$\tau > \tau_+$ regions. 
The adiabatic initial conditions are $\delta(k,0) = -2\phi(k, 0)$ and $\delta'(k,0) = 0$ 
for modes that entered the horizon before the FOPT.
We introduce
%
% \begin{align}
%   \delta(k, \tau) &= \mathcal{R}(k) \times
%   \begin{cases}
%     \mathcal{G}_- (x) \, , & \tau < \tau_- \, ,
%     \\
%     \mathcal{G}_+ (x) \, , & \tau > \tau_+ \, ,
%   \end{cases}
% \end{align}
%
\begin{align}
  \delta(k, \tau) &= \mathcal{R}(k) \mathcal G_\mp(x) \, , & \tau &\lessgtr \tau_\mp \, ,
\end{align}
where the $\mathcal{G}_\mp$ depend only on the dimensionless $x$ as
\begin{align}\label{eq:Fpm}
  \!\! \mathcal{G}_-(x)&= 4 \cos x + 8 \, \frac{(1 - x^2 )\sin x  - x \cos x}{x^3} \, ,
  \\ \nonumber
  \!\! \mathcal{G}_+(x)&= c_1 + c_2 \ln x + 6 {\rm Ci} \left( x \right) + 
  6 \frac{(1 - x^2)\sin x  - x \cos x}{x^3} \, ,
\end{align}
where $\text{Ci}$ is the cosine integral function.
The two coefficients $c_1, c_2$ are fixed by matching $\mathcal{G}_-$ to 
$\mathcal{G}_+$ across the FOPT $\tau_- \leq \tau \leq \tau_+$.

During the transition, the BFDM equation of state varies as in~\eqref{eq:modelw}, 
or in terms of the conformal time 
\begin{align}\label{eq:EOSconformal}
  w(\tau) \simeq \frac{1}{3} \exp \left\{ - \exp \left[\frac{\beta}{2H_*}
  \left(\frac{\tau^2}{\tau_*^2} - 1\right) \right] \right\} \, , 
\end{align}
and using $w = p/\rho$ we obtain ${\delta p}/{\delta \rho} = w + {\rho w'}/{\rho'}$,
where $w' \simeq -1/3/(\tau_+ - \tau_-)$. 
In the $\tau_+ \to \tau_-$ limit, $w'$ becomes large.
The derivative of the energy density follows the continuity equation,
$\rho' = -3(1+w) \rho a'/a$, leading to
\begin{align}\label{eq:dpdrho}
  \frac{\delta p}{\delta \rho} = w - \frac{w'}{3 (1 + w) a'/a} \, .
\end{align}
The continuity equation implies that $\rho$ is preserved in the instantaneous FOPT limit. 
In order for dark matter particles to acquire a large mass in the new phase, they must interact with the expanding bubble wall.
Eq.~\eqref{eq:dpdrho} is valid if the dark sector is secluded, and the bubbles transfer negligible amount of energy or they themselves are part of the dark sector.
These constitute the blast-frozen condition investigated in this work.
A promising class of models to realize the above condition are hidden Yang-Mills theories without light quark where a FOPT is known to happen~\cite{Pisarski:1983ms, Schwaller:2015tja, Halverson:2020xpg}. 
Lattice simulations show that glueball masses are well above the confinement scale, fulfilling the blast-freezing condition~\cite{Panero:2009tv, Falkowski:2009yz, Soni:2016gzf, Curtin:2022tou}.

In the alternative case where the energy density $\rho$ experiences a sudden change, the effect can be modeled by adding a source term to the continuity equation, whose form depends on the details of the dark sector model. 
This will modify the form of $\delta \rho/\delta p$ in Eq.~\eqref{eq:dpdrho}.
However, the source term must vanish on superhorizon scales, which is important for deriving the adiabatic initial conditions for dark matter perturbations (see below).

\begin{figure}
  \centering
  \includegraphics[width=1\columnwidth]{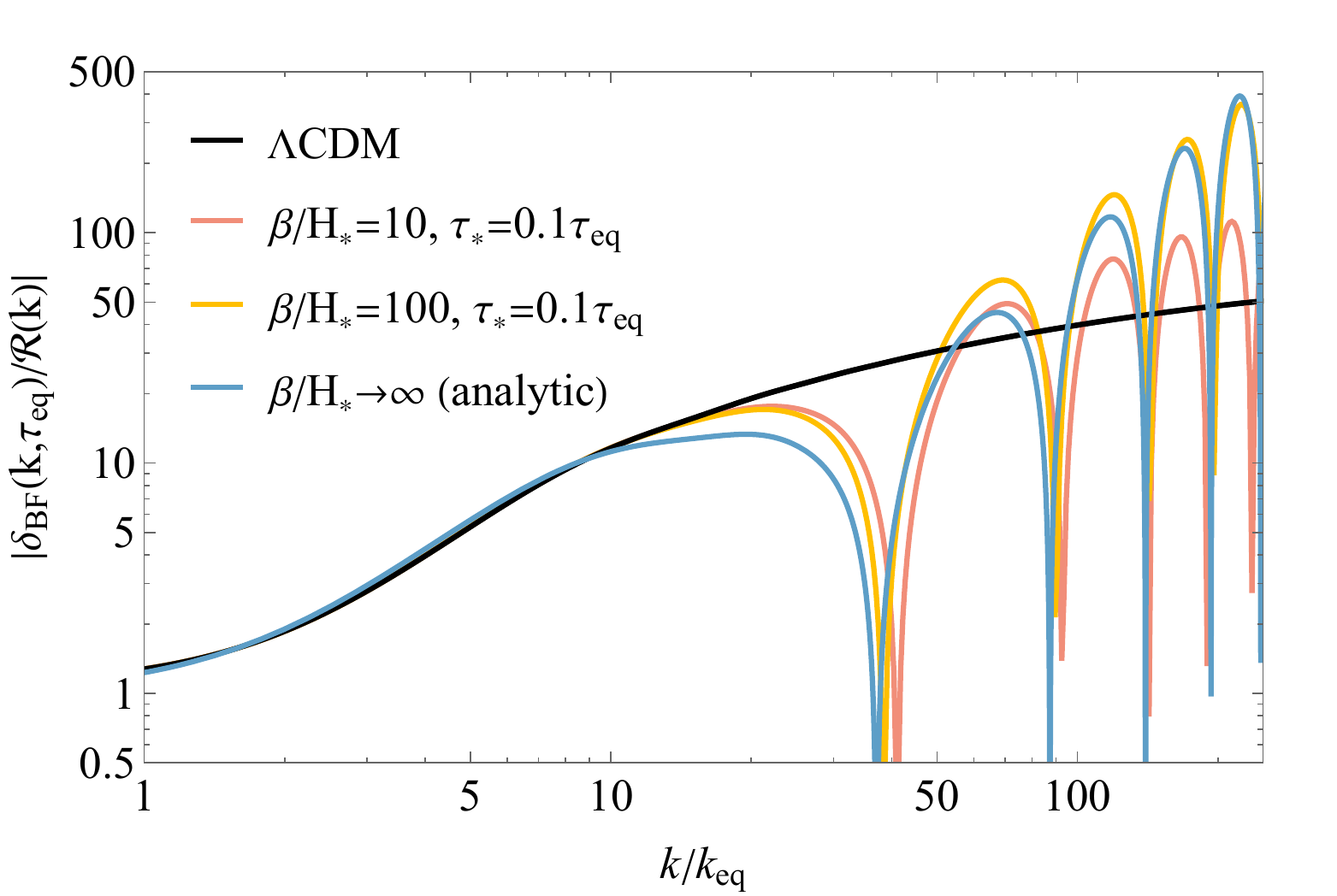}
  \caption{DM density perturbations evolved to the time of MRE. 
  The $\Lambda$CDM model predicts the smooth black curve, whereas the BFDM features 
  oscillations in momentum space shown by the red and yellow curves, corresponding to
  $\beta/H_*=10$ and 100, respectively.
  We assume the FOPT occurs at conformal time $\tau_*=\tau_{\rm eq}/10$ and BFDM
  comprises all of DM in the universe.
  The analytic solution in blue follows from~\eqref{eq:G+case1} and is derived
  in radiation dominated universe.} \label{fig:2}
\end{figure}

Working in the instantaneous FOPT limit with $\tau_+ \to \tau_-$ and $\beta/H_* \gg 1$, 
allows for an analytic solution of the density perturbation equations.
Here, the $w'$ term dominates in the first equation of~\eqref{eq:DMPert1}, 
which becomes
\begin{equation}
  \frac{\text{d}\delta}{\text{d}w} \simeq \frac{\delta}{1 + w} \, ,
\end{equation}
where we assume that $w$ drops monotonically and use it as the effective 
``time'' during the FOPT. 
Integrating both sides leads to the approximate solution for 
$\tau_- \leq \tau \leq \tau_+$ and the following matching condition for $\delta(\tau_\pm)$,
\begin{align}\label{eq:BC1}
  \delta(\tau)   &= \frac{3}{4} \left(1 + w(\tau) \right) \delta(\tau_-) \, , 
  & 
  \delta(\tau_+) &= \frac{3}{4} \delta(\tau_-) \, ,
\end{align}
where we used $w(\tau_-) = 1/3$ and $w(\tau_+) = 0$ in Eq.~\eqref{eq:genericw}.
Remarkably, for superhorizon modes, the above matching conditions allow the primordial perturbations to remain adiabatic~\cite{Baumann:2022mni, Langlois:2003fq} as
dark particles transition from radiation to matter through the FOPT.
In the End Matter~\hyperref[ed2]{II}, we comment on an earlier work~\cite{Das:2018ons} that considered a different matching condition.

With the boundary conditions $w'(\tau_-) = w'(\tau_+) = 0$ 
from~\eqref{eq:genericw}, the first equation of \eqref{eq:DMPert1} implies
\begin{equation}\label{eq:BC2}
  \delta'(\tau_+) - \delta'(\tau_-) \simeq \frac{4}{3} \theta(\tau_+) - \theta(\tau_-) \, ,
\end{equation}
where we neglect the gravitational potential $\phi$ compared to $\delta$ in 
radiation dominated universe.
The right-hand side of~\eqref{eq:BC2} can be solved using the second equation 
of~\eqref{eq:DMPert1}, which in the large $w'$ limit reads
\begin{equation}
  \frac{\text{d} \theta}{\text{d} w} \simeq - \frac{\theta}{1 + w} - 
  \frac{k^2 \tau_*}{4(1 + w)} \delta(\tau_-) \, .
\end{equation}
The solution for $\tau_- < \tau < \tau_+$ is
\begin{equation}
  \theta(w) = \frac{d}{1+w} - \frac{w}{4(1+w)} k^2\tau_* \delta(\tau_-) \, ,
\end{equation}
where $d$ is a constant of integration that cancels away in~\eqref{eq:BC2}, such that
\begin{equation}\label{eq:BC2b}
  \delta'(\tau_+) - \delta'(\tau_-) \simeq -\frac{1}{12} k^2\tau_* \delta(\tau_-) \, .
\end{equation}

Applying the two matching conditions in~\eqref{eq:BC1} and \eqref{eq:BC2b} to the 
DM density perturbation solutions in Eq.~\eqref{eq:Fpm} fixes $c_{1, 2}$.
The corresponding solution for perturbations after the FOPT, but before MRE, is then
\begin{align}\label{eq:G+case1}
  \mathcal{G}_+ &= \frac{6}{x^3} \left[-x\cos x + (1-x^2)\sin x\right]  + 6 \left[ 
  \text{Ci}(x) - \text{Ci}(x_*)\right]  \nonumber
  \\
  &+ 3 \cos x_* - \frac{1}{x_*^3} \log(x/x_*) \left[ x_* \left( x_*^4 + 6 x_*^2 - 6 \right) 
  \cos x_* \rule{0mm}{4mm}\right. \nonumber 
  \\
  &+\left. 2 (x_*^4 - x_*^2 +3) \sin x_* \rule{0mm}{4mm}\right] \, ,
\end{align}
where $x_* = k\tau_*/\sqrt{3}$.
Sub-horizon modes, which enter the horizon well before the FOPT with $x > x_*\gg 1$, 
take the asymptotic form
\begin{equation}\label{eq:SolutionCase1}
  \delta (x) \simeq - x_*^2 \cos x_* \log\left( \frac{x}{x_*} \right)\mathcal{R}(k) \, .
\end{equation}

\begin{figure}
  \centering
  \includegraphics[width=0.95\columnwidth]{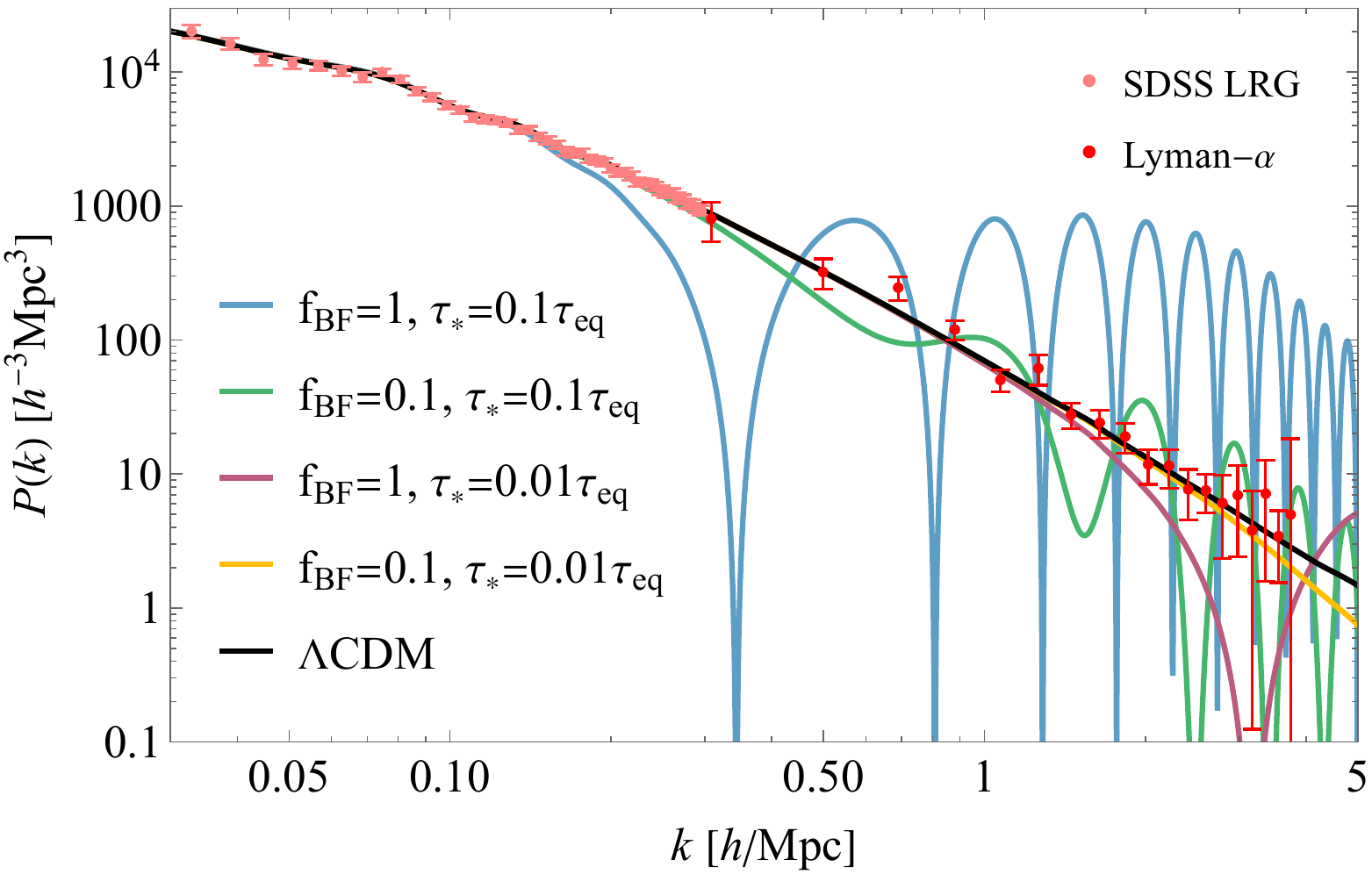}
  \caption{
  The total matter power spectrum, linearly evolved until today, for BFDM scenarios 
  with various combinations of $f_{\rm BF}$ and $\tau_*$ parameters (colored curves) 
  and fixed $\beta/H_* = 100$.
  The $\Lambda$CDM model is shown in black, together with the data points from the
  observations of SDSS LRG in pink and BOSS Lyman-$\alpha$ in red~\cite{Chabanier:2019eai}. 
  %
  %The agreement with the $\Lambda$CDM model allows the $P(k)$ spectrum to play an instrumental role constraining the BFDM.
  } \label{fig:3}
\end{figure}

In addition to the logarithmic growth for DM density perturbations, expected 
in radiation dominated universe, the BFDM solution in~\eqref{eq:SolutionCase1} 
features a prefactor $x_*^2 \cos x_*$. 
This leads to oscillations of $|\delta|$ in the wave number $k$ space 
with peaks and zeroes.
The peaks are located at $x_* \simeq \text{integers}\times \pi$ and they
grow with $k$, while the zeroes occur at $x_* \simeq \text{odd integers}\times \pi/2$.
These features are demonstrated by the red curve in FIG.~\ref{fig:2}, which plots 
Eq.~\eqref{eq:G+case1}, the density perturbation for the BFDM at the 
time of MRE as a function of $k$.~\footnote{The subsequent growth of $\delta$ between MRE and today is almost linear in the scale factor $a$ thus the oscillating features will be preserved until observations are made. See Fig.~\ref{fig:3}.}
In the smaller $k$ region, perturbation modes enter the horizon after the 
FOPT (with $x_* \ll 1$) and take the usual cold DM (CDM) form
$\delta_+ (x) = - (6\log x + 6 \gamma - 3)\mathcal{R}(k)$ at late times 
($x\gg 1$), where the coefficient of $\mathcal{R}(k) \log x$ is $k$-independent.

The physics behind such modulating behavior can be understood intuitively.
Prior to the FOPT, DM is massless and its fluid exerts pressure against gravity.
Its perturbations therefore oscillate with time, as shown by $\mathcal{G}_-$ in~\eqref{eq:Fpm}. 
For subhorizon modes with $x_* \gg 1$, the value of $\delta$ immediately before FOPT is 
proportional to $\cos x_*$ that oscillates with $k$.
This sets the initial value for the logarithmic growth after the FOPT. 
When DM undergoes blast freezing, the memory of its past as radiation is carried along.
The additional $x_*^2$ factor is the consequence of the matching condition in~\eqref{eq:BC2b}
and strongly amplifies the perturbations for certain large $k$ (on smaller scales).

{\bf Modulated matter power spectrum.} 
Let us quantify the implications of oscillating density perturbations of the blast-frozen
DM in cosmology. 
The most direct and important experimental probe is the DM two-point power spectrum, observed 
for large and small scale structures.
We work with a two-component DM, where the total matter power spectrum can be written as 
\begin{equation}
  P(k) = P_{\rm \Lambda CDM}(k) \left|\frac{f_{\rm BF}\delta_{\rm BF} +
  (1 - f_{\rm BF}) \delta_{\rm CDM}}{\delta_{\rm \Lambda CDM}}\right|^2 \, . 
\end{equation}
The BFDM comprises a fraction $f_{\rm BF}$ of today's DM relic
density and $\delta_{\rm BF}$ is its density perturbation calculated above.
The rest of the DM fills the remaining fraction $1 - f_{\rm BF}$ and consists of
the regular cold DM with perturbations $\delta_{\rm CDM}$.
The total matter power spectrum is normalized to $\Lambda$CDM 
with $\delta_{\rm \Lambda CDM}$ perturbations.
For $0 < f_{\rm BF} < 1$, the evolution of $\delta_{\rm CDM}$ gets 
gravitationally affected by the oscillating blast-frozen component and deviates 
from $\delta_{\rm \Lambda CDM}$, and vice versa.
This interplay is more profound around and after the MRE.
The analytic solution for $\delta_{\rm BF}$ found above is independent of 
$f_{\rm BF}$, because it was derived for a radiation dominated universe neglecting back-reactions from CDM.

To account for $f_{\rm BF}$ and for the evolution of perturbations when the universe 
approaches the matter dominated era, we numerically solve a coupled system 
of differential equations, including Eq.~\eqref{eq:DMPert1}, along with the density 
perturbation equations for radiation and CDM, and the Einstein equation.
The resulting matter power spectra for BFDM are depicted in FIG.~\ref{fig:3} for
various combinations of $f_{\rm BF}$ and $\tau_*$, while holding $\beta/H_* = 100$.
All the density perturbations are linearly extrapolated to today in order to
compare with observations.

\begin{figure}
  \centering
  \includegraphics[width=1\columnwidth]{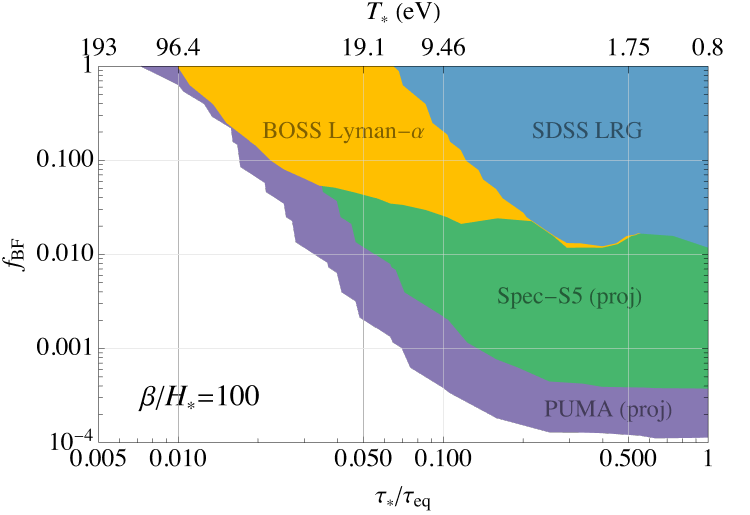}
  \caption{
  Constraints on BFDM in the $f_{\rm BF}$ versus $\tau_*/\tau_{\rm eq}$ parameter space 
  based on the matter power spectrum measurements made by SDSS (blue shaded region) and 
  BOSS (yellow shaded region) for $\beta/H_* = 100$. 
  The upcoming cosmological experiments based on spectroscopic and 21-cm surveys 
  can greatly expand the sensitivity in the BFDM parameter space, as shown by 
  the green and purple shaded regions, respectively.} \label{fig:4}
\end{figure}

In the presence of BFDM, the total $P(k)$ spectrum inherits the $\delta_{BF}$
oscillations.
As shown in FIG.~\ref{fig:3} the $P(k)$ oscillation peaks can 
exceed the $\Lambda$CDM counterparts by far.
This tends to run in conflict with the existing data~\cite{Reid:2009xm, BOSS:2012tuc},
with its significance controlled by $f_{\rm BF}$ and $\tau_*$.
A smaller $f_{\rm BF}$ reduces the oscillation amplitude, whereas a smaller 
$\tau_*$ implies an earlier FOPT that pushes the onset of oscillations to 
higher $k$.
The reference curve for $\Lambda$CDM is produced using CLASS~\cite{Blas:2011rf} 
with fiducial cosmological parameters $h = 0.678$, $\Omega_{\rm DM}h^2 = 0.12$, 
$\Omega_b h^2 = 0.022$, $A_s = 2.101 \times 10^{-9}$, $n_s = 0.966$ and 
$\tau_{\rm reio} = 0.054$.

In FIG.~\ref{fig:4} we scan over the $f_{\rm BF}$ versus $\tau_*$ parameter space 
and confront the predicted matter power spectrum to the corresponding observations 
of Luminous Red Galaxies (LRG) by SDSS DR7 and Lyman-$\alpha$ forest by BOSS DR9, 
which exclude the blue and yellow shaded regions, respectively. 
The striking oscillations of the predicted $P(k)$ spectrum allow robust
constraints to be set on BFDM.
We find that the BFDM may comprise 100\% of DM in the universe only if the FOPT 
occurs at a sufficiently early conformal time $\tau_* \lesssim 0.01 \, \tau_{\rm eq}$, 
corresponding to the photon temperature $T_* \gtrsim 96 \, $eV and beyond the 
smallest structure probed by the Lyman-$\alpha$ forest.
Conversely, if the FOPT takes place at $\tau_* \gtrsim 0.05 \, \tau_{\rm eq}$, 
the existing cosmological data require the BFDM fraction not to exceed a few 
percent of the total DM.

We also show the projected sensitivity of future cosmological surveys, including the 
Stage-5 Spectroscopy (Spec-S5) and 21-cm mapping array 
(PUMA)~\cite{Sailer:2021yzm, Schlegel2023}.
These promise to test the primordial matter power spectrum with higher precision on 
both large and small scales.
They can probe the presence of BFDM with a fraction of total DM as small as 
$\sim 10^{-4}$, or the FOPT as early as several hundred eV.
The corresponding coverages are shown by the green and purple shaded regions in 
FIG.~\ref{fig:4}.
For cases with smaller $\beta$ we find that the constraints and projections remain 
similar, but the PUMA coverage extends to higher $T_*$, up to several hundred eV.

Prior to the FOPT, dark matter particles were relativistic and contribute to the radiation energy density of the universe. 
The contribution can be obtained by extrapolating back in time from MRE, when the photon temperature is $T_{\rm eq} \simeq 0.8 \,$eV.
Around and before the FOPT, the energy density ratio of dark radiation to photons and neutrinos is $T_{\rm eq} f_{\rm BF}/T_*$. 
It translates into $\Delta N_{\rm eff} \simeq 0.14 f_{\rm BF} (T_{\rm eq}/T_*)$ which, given the existing constraints shown in Fig.~\ref{fig:4}, is well below than the limit set by the big-bang nucleosynthesis~\cite{Pitrou:2018cgg}.

{\bf Conclusion and Outlook.}
To summarize, we explore the cosmological implications of an early universe FOPT, 
which serves the well-motivated role of generating the DM mass.
We consider the blast freezing scenario where the DM's equation of state makes an 
abrupt change from 1/3 to 0 and point out the smoking-gun signature of sharp modulations 
in the primordial matter power spectrum (presented as $P(k)$ in FIG.~\ref{fig:3}). 
The latter act as seeds for the subsequent structure formation.
The existing cosmological data are sensitive to BFDM and the FOPT with nucleation 
temperature $T_* \lesssim 100$ eV and restricts the BFDM to comprise less than a few 
percent of total the DM if the FOPT occurs near the time of MRE. 
Further cosmological surveys will extend the probe of $T_*$ up to $\sim$ keV scale.
Interestingly, the probe using DM power spectrum is complementary to the search for 
GWs, which are most sensitive to FOPTs before the onset of big-bang nucleosynthesis.

With the upcoming wave of cosmological data in the next decade~\cite{Chou:2022luk} there lies
the exciting opportunity to uncover physics beyond the $\Lambda$CDM model.
It would be useful to go beyond the simple 1D $P(k)$ fit done here and confront the BFDM 
with a global analysis, including all cosmological tracers, with proper forward 
modeling using e.g. the EFTofLSS developed in~\cite{Baumann:2010tm, Carrasco:2012cv, 
Bottaro:2023wkd}.
A more thorough analysis may take into account the spatial dependence of the DM 
equation of state $w$, which has been neglected in Eq.~\eqref{eq:EOSconformal} in 
the spirit of prompt phase transition ($\beta/H_*\gg1$).
For $\beta \sim H_*$, random bubble nucleation would imply that the blast freezing
happens at different $\tau_*(\vec{x})$ throughout the universe and contribute 
to additional $k$ dependence in the final $P(k)$ spectrum.
Another natural generalization is to consider the finite DM mass 
after the FOPT, where the matter power spectrum is further processed by a nonzero DM 
velocity dispersion, akin to the effect of warm DM.

{\bf Acknowledgements.}
We thank Anže Slosar and Mark Wise for useful discussions and correspondence and Noah
Sailer for sharing the projection data from~\cite{Sailer:2021yzm}. 
MN is supported by the Slovenian Research Agency under the research core funding 
No.~P1-0035 and in part by the research grants N1-0253, J1-4389 and J1-60026.
YZ is supported by a Subatomic Physics Discovery Grant (individual) from the Natural 
Sciences and Engineering Research Council of Canada.
MN and YZ are supported by an International Research Seed Grant awarded by Carleton
University.
MN would like to thank the particle theory group at Carleton University for 
hospitality during the stay when this work was initiated.

\section{End Matter}

\subsection{I. Volume fraction of the false vacuum}\label{ed1}

Here we provide additional technical details on the derivation of Eq.~\eqref{eq:modelw}, 
which is key to our analysis.
We recap the time evolution of the fractional volume ${\cal T}(t)$ of 
the true vacuum (TV) in an expanding FLRW universe~\cite{Guth:1982pn}.
Below the critical temperature $T_c$, corresponding to the critical time $t_c$, 
tunneling is allowed, the first bubbles appear and expand to fill out the TV volume.
In vacuum, the bubble wall would quickly reach the speed of light~\cite{Coleman:1977py} 
and obey $\text{d} s^2 = \text{d} t^2 - a(t)^2 \text{d} r^2 = 0$, but when interacting 
with the plasma, its wall velocity is given by $v_w$.
Thus the radius $R$ and the volume $V$, swept out by the bubble wall nucleated at 
$t_1 < t$, are given by
\begin{align}
  R(t_1, t) &= v_w \int_{t_1}^t \frac{\text{d} t'}{a(t')} \, ,
  &
  V(t_1, t) &= \frac{4 \pi}{3} R(t_1, t)^3 \, .
\end{align}
The number of bubbles produced in a time interval $\text{d} t$ is given by 
$\text{d}N = \gamma(t) \text{d} t \, a(t)^3 \, {\cal F}(t)$, where ${\cal F}$ 
is the fractional volume of the false vacuum (FV) upon which bubbles can form, and
${\cal F} + {\cal T} = 1$.
The total volume of TV within all the bubbles nucleated after $t_c$ is then
\begin{align} \label{eq:TVFracVol}
  {\cal T}(t) &= 1 - {\cal F}(t) = \int_{t_c}^t \text{d} t' \gamma(t') 
  a(t')^3 V(t', t) {\cal F}(t') \, ,
\end{align}
which turns into an iteration equation that is solved by
\begin{align} \label{eq:FVFracVol}
  {\cal F}(t) &= \exp \left( \int_{t_c}^t \text{d} t' \gamma(t') a(t')^3 
  V(t', t) \right) \, .
\end{align}
In radiation domination, the Hubble parameter is $H \propto T^2 \propto a^{-2}$, 
such that $H a^2$ remains constant and we can evaluate the $V(t', t)$ as
\begin{align}
  V &= \frac{4 \pi}{3} \left( \int_{t'}^t \frac{\text{d} t'' v_w}{a(t'')} \right)^3 =
  \frac{4 \pi}{3} \left(v_w \frac{ a(t) - a(t')}{H_* a_*^2} \right)^3 \, , 
\end{align}
where we switched the integration variable from $t$ to $a(t)$ and took out the 
$H a^2 = H_* a_*^2$ constant term.
In radiation domination, $a(t) \propto \sqrt t$, which results in a power-law growth 
with time.
Conversely, the $\gamma(t) \sim A \exp{(-B(t))}$ term in~\eqref{eq:FVFracVol} has an
exponential $t$ dependence that dominates~\cite{Enqvist:1991xw} the integral 
in~\eqref{eq:FVFracVol}.
We can expand the integrand around $t_*$ by introducing the usual~\cite{Caprini:2019egz} 
$\beta = \text{d}(\ln \gamma(t))/\text{d} t  \simeq -\text{d} B/\text{d} t$
parameter at the time of nucleation, such that
\begin{align} \label{eq:FVFracBeta}
  {\cal F}(t) &= \exp \left( -\int_{t_c}^t \text{d} t' \, g(t') e^{-B(t')} \right)
  \\
  &\simeq \exp \left( -g(t_*) e^{-B(t_*)} \int_{t_c}^t \text{d} t' \, e^{\beta(t'-t_*)} \right) \, ,  
\end{align}
where $g(t)$ describes a generic non-exponential behaviour of $A(t) a(t)^3 V(t_1,t)$.
We then define the nucleation time as ${\cal F}(t_*) = 1/e$~\cite{Enqvist:1991xw}, 
such that
\begin{align} \label{eq:FVFracBeta2}
  {\cal F}(t) \simeq \exp \left( - e^{\beta (t - t_*)} \right) \, ,
\end{align}
where we dropped an exponentially suppressed $\exp(-\beta(t_* - t_c))$ term.
Due to the exponential temperature dependence in the nucleation rate $\gamma$, our 
nucleation temperature $T_*$ is numerically very close to the usual definition of 
the nucleation temperature $T_n$ given by $\int_{T_n}^\infty \gamma \text{d}T/(T H^3) = 1$,
or approximately $\gamma/H^4|_{T_n} = 1$~\cite{Quiros:1999jp}.

\subsection{II. Comment on other phase transition scenarios}\label{ed2}

To derive the matching conditions in Eqs.~\eqref{eq:BC1} and \eqref{eq:BC2b}, we assume that the energy of a secluded dark sector evolves continuously during the FOPT.
In the instantaneous phase transition limit, the dark sound speed $\delta p/\delta \rho$ in Eq.~\eqref{eq:dpdrho} blows up along with the $w'$ term.
As emphasized in the main text, this is crucial for the perturbations to respect the adiabatic initial condition~\cite{Baumann:2022mni, Langlois:2003fq} 
\begin{equation}\label{eq:BCCase1app}
\delta(\tau_+) = \frac{3}{4} \delta(\tau_-) = \frac{3}{4} \delta_\gamma \ , 
\end{equation}
before and after the FOPT, where $\delta_\gamma$ denotes the primordial perturbation of radiation species (photons).

Next, we comment on a different set matching conditions used in an earlier work~\cite{Das:2018ons}
\begin{align}\label{eq:BCCase2}
  \delta(\tau_+) &= \delta(\tau_-)\ , \quad \delta'(\tau_+) = \delta'(\tau_-) \ ,
\end{align}
which were claimed to be derived assuming the continuity of the stress-energy tensor during the phase transition, in contradiction to our derivation  in the main text.
Eq.~\eqref{eq:BCCase2} applies to all density perturbation modes, including those that are still superhorizon at the time of FOPT but later enter the horizon to form the large scale structures of the universe.

We point out that Eq.~\eqref{eq:BCCase2} suffers from a severe problem for leading to too large isocurvature perturbations.
Indeed, before the phase transition, the dark sector has $w=1/3$ and the primordial perturbations are equal to those of the photons, $\delta(\tau_-) = \delta_\gamma$.
Eq.~\eqref{eq:BCCase2} then implies that $\delta(\tau_+) = \delta_\gamma$ after the phase transition when dark matter already become non-relativistic with $w=0$.
This is in sharp contradiction to the adiabatic initial condition derived in Eq.~\eqref{eq:BC1} and \eqref{eq:BCCase1app}, and the resulting cosmology will be inconsistent with the existing cosmic microwave background observations~\cite{Planck:2018jri}.
Therefore, we conclude that the scenario explored in~\cite{Das:2018ons} is not realistic.

\bibliographystyle{apsrev4-1}
\bibliography{refs}{}
\end{document}